\documentclass[pre,amsmath,amssymb,floatfix,twocolumn]{revtex4}
\usepackage{graphicx}
\usepackage{float}
\usepackage{bm}
\usepackage{color}
\usepackage[utf8]{inputenc}

\begin{document}

\title{The nature of the ordered phase of the confined self-assembled rigid rod model}
\author{N. G. Almarza}
\affiliation{Instituto de Qu{\'\i}mica F{\'\i}sica Rocasolano, CSIC, Serrano 119, E-28006 Madrid, Spain }
\author{J. M. Tavares}
\affiliation{Centro de F\'{\i}sica Te\'orica e Computacional, Universidade de Lisboa, Avenida Professor Gama Pinto 2,
P-1649-003 Lisbon, Portugal}
\affiliation{Instituto Superior de Engenharia de Lisboa, Rua Conselheiro Em\'{\i}dio Navarro 1, 
P-1950-062 Lisbon, Portugal}
\author{M. M. Telo da Gama}
\affiliation{Centro de F\'{\i}sica Te\'orica e Computacional, Universidade de Lisboa, Avenida Professor Gama Pinto 2,
P-1649-003 Lisbon, Portugal}
\affiliation{Departamento de F\'{\i}sica, Faculdade de Ci\^encias, Universidade de Lisboa, Campo Grande,
P-1749-016 Lisbon, Portugal}

\begin{abstract}
We investigate the nature of the ordered phase and the orientational correlations between adjacent layers of the confined three-dimensional self-assembled rigid rod model, on the 
cubic lattice. We find that the ordered phase at finite temperatures becomes uniaxial in the thermodynamic limit, by contrast to the ground state (partial) order where the orientation of the uncorrelated layers is perpendicular to one of the three lattice directions. The increase of the orientational correlation between layers as the number of layers increases suggests that the unconfined model may also exhibit uniaxial ordering at finite temperatures. 

\end{abstract}
\maketitle

\section{Introduction}
\label{sec.intro}

State of the art techniques for synthesizing colloids monodisperse in shape and size allow their collective behavior to be investigated \cite{Pawar}. The new particles may become the `molecules' of new materials if they can be tailored and assembled into useful structures \cite{Glotzer}. In fact, the possibility of particle decoration (through, e.g., glancing angle deposition, templating, or lithography) produces colloids with pre-determined surface patterns (patches). Patches yield new features such as anisotropic interactions, control of the valence, and the formation of permanent electrical dipoles, paving the way for the rational development of novel self-assembled materials (e.g., super-molecules) with highly tunable mechanical, optical, and thermal properties \cite{Pawar,Sciortino}. 

Self-assembly has been exploited theoretically for a primitive model of patchy colloids and state of the art simulation studies revealed how the number, type, and distribution of the patches determine the self-assembled structures. In systems with two bonding sites per particle, only (polydisperse) linear chains form and there is no liquid-vapor phase transition \cite{Sciortino}. If the linear chains are stiff they will undergo an ordering transition, at fixed concentration, as the temperature decreases. The minimal model of this transition considers the effects of the equilibrium polydispersity and the polymerization process of the rods. In this context, we proposed a model of self-assembled rigid rods (SARR), composed of monomers with two bonding sites that polymerize reversibly into polydisperse chains \cite{Tavares2009a} and carried out extensive Monte Carlo simulations to investigate the nature of the ordering transition on the square and triangular lattices \cite{Almarza2010,Almarza2011}. The polydisperse rods undergo a continuous ordering transition that was found to be in the two-dimensional (2D) Potts q=2 (Ising) and q=3 universality classes, respectively, as in similar models where the rods are monodisperse \cite{Matoz2008b}. These findings refute previous claims, based on Canonical Monte Carlo simulations, that equilibrium polydispersity and the statistical ensemble change the criticality of these models to random percolation \cite{Lopez,Comment,Reply}.

The nature of the ordering transition of the three-dimensional (3D) SARR model  on
the simple cubic lattice is much more difficult to establish. The model consists of particles with two patches aligned along $\pm \hat{\alpha}$, where $\hat{\alpha}$ represents one of the three lattice directions ($x$,$y$,$z$). Particles on nearest-neighbor (NN) lattice sites ${\bf r}_i$ and ${\bf r}_i + \hat{\alpha}$ interact attractively with energy $-\epsilon$ if their patches are aligned along $\hat {\alpha}$.
Monte Carlo Simulations using efficient algorithms suggest that the ordered phase (below the transition temperature) exhibits a bias towards uniaxial behavior (i.e. the system exhibits a tendency to align different layers, by contrast to the ground state partial order). This tendency is observed only when the system size, defined by $L$ with $L^3$ the number of sites considered in the simulation, is sufficiently large at temperatures that are not too low, $T>>0$. Despite the use of efficient cluster algorithms we have not been able to establish the nature of the ordered phase as the system sizes required to observe uniaxial behavior increase rapidly as the temperature decreases, as discussed below. 

Here we consider the confined 3D SARR model as a first step towards elucidating the nature of the ordered transition on the cubic lattice. We investigate the nature of the ordered phase and the orientational correlations between adjacent layers of the confined model, on the cubic lattice, and find that the ordered phase at finite temperatures becomes uniaxial in the thermodynamic limit, by contrast to the ground state order where the orientation of the uncorrelated layers is perpendicular to one of the three lattice directions. In addition, we find that the orientational correlation between layers increases as the number of layers increases from two to three suggesting that the unconfined model may also exhibit uniaxial order at finite temperatures.

The paper is arranged as follows: In section II we describe the ground state and the simulation methods used to analyze the 
3D SARR model at finite temperatures while in section III we present the simulation results for the ordering transition and the order parameters. In section IV we introduce the confined SARR model. In section V we present the simulation results for the order parameters and the correlations between adjacent layers, for models with two and three layers. We conclude in section VI with a discussion of the results. 

\section{Three-dimensional SARR model}

\subsection{Ground state}
\label{ssec:gs}

In the full lattice limit every site is occupied by one particle aligned in one of the three lattice directions. In the ground state, the SARR model exhibits partial order: One lattice direction (say, $z$) is suppressed, with the uncoupled layers aligned in one of the remaining lattice directions ($x$ or $y$). The ground state potential energy is then $U=-N\epsilon$, where $N$ is the number of lattice sites, with degeneracy:
\begin{equation}
\omega_{GS} = 3 \times ( 2^L-1 ).
\end{equation}
The entropy per site vanishes in the thermodynamic limit.

\subsection{Simulation procedures}
\label{sec:sim}

It has been shown that the full-lattice 2D SARR model on the square lattice can be mapped on the 2D Ising model \cite{Almarza2010}. This mapping allows the use of cluster algorithms developed for Potts models\cite{Swendsen,Landau-Binder} to enhance the efficiency of the Monte Carlo simulations. In what follows we describe how the Swendsen-Wang algorithm may be adapted to the SARR model on lattices where the mapping does not exist. We recall that the 3D SARR model on the cubic lattice or the 2D SARR model on the triangular lattice cannot be mapped on Ising or Potts models \cite{Almarza2011}. We can, however, develop cluster algorithms based on the layer structure of ground state of the 3D SARR model and the mapping of the 2D SARR model on the square lattice.

\subsubsection{Cluster sampling}

The cluster algorithm samples at each MC step a subset of all sites as described next. One of the three lattice directions is chosen at random, say ${\hat z}$. Then the sites oriented along ${\hat z}$ are blocked, i.e. their state is frozen during the MC step. Sites with orientations ${\hat x}$ or ${\hat y}$ are {\it active}, and their states may change during the MC step.
Given the NN character of the Hamiltonian the procedure may be (and it is) applied to all active sites in one MC step. 
For simplicity, however, we consider one layer, i.e. all sites, $i$, with $z_i=z_0$. The procedure starts by checking the 
links between pairs of NN active sites (not to be confused with {\it bonds} of the original model). Two NN active sites with the same orientation are linked with probability $B=1-\exp(-\beta \epsilon/2)$, where $\beta \equiv (k_BT)^{-1}$. Links cannot be formed between pairs of NN active sites with different orientations. We define clusters of active sites based on the links generated in the previous step. The new configuration is obtained by choosing, independently, for each cluster a new in-plane orientation. The probability of the new cluster orientation, ($\hat{x}$ and $\hat{y}$), is given by:
\begin{equation}
A_k({\hat \alpha}) \propto \exp \left[ - \frac{\beta \epsilon}{2} n_k({\hat \alpha}) \right];
\end{equation}
where $n_k({\hat \alpha})$ is the number of patches of the cluster $k$ that point to blocked sites when the active sites are 
oriented along ${\hat \alpha}$. 

The procedure is validated using the plaquette formalism\cite{Almarza2010} that maps the model with blocked sites orientated along $\hat{z}$ to a Potts model \cite{Wu} in an external field. 
For a system with $L$ layers ($z=1,2,\cdots,L$) the intralayer potential energy can be written as:
\begin{equation}
{\cal U} = - \epsilon \sum_{<ij>}\delta({\hat \alpha}_i,{\hat \alpha}_j)  \delta (\hat{\alpha}_i,\hat{r}_{ij});
\end{equation}
where $<ij>$ runs over the NN active sites in one layer, $\hat{\alpha}_i$ is the orientation of site $i$, and $\delta({\hat \alpha},{\hat \alpha'})$ is one if $|{\hat \alpha} \cdot {\hat \alpha'}|= 1$, and zero otherwise. This intralayer Hamiltonian can be mapped to a $q=2$ Potts model\cite{Wu} with blocked sites on the square lattice. Using the plaquette formalism \cite{Almarza2010} we find:
\begin{equation}
{\cal U} = - K \sum_{<ij>} \delta ({\hat \alpha}_i,{\hat \alpha_j}) 
- \sum_{<ik]} \left[ K_0 + K_1 \delta({\hat \alpha}_i,{\hat r}_{ik}) \right].
\label{potts}
\end{equation}
where the subindex $<ik]$ runs over pairs of NN sites on the layer with $i$ an active site and $k$ a passive one. In Eq. (\ref{potts}) $K$ is the coupling constant, while $K_0$ and $K_1$ describe the interactions between active sites and the blocked ones (this may be viewed as the interaction of an external field with the active sites). $K$, $K_0$, and $K_1$ are given in terms of the energy of the patchy model by: $K=\epsilon/2$, $K_0=\epsilon/4$, and $K_1=-\epsilon/2$. As expected the interaction energy of the active sites pointing to  blocked ones is unfavorable.  
It is now straightforward to implement the cluster algorithm described above. The Potts coupling defines linking criterion between active sites according to the Swendsen-Wang rules \cite{Swendsen} while the single-particle interactions are taken 
into account by considering the effect of an external field \cite{Landau-Binder}.

\subsubsection{Sublattice sampling}

In addition to the cluster moves we implemented sublattice single-particle moves. We consider systems with $L$ even, and divide the sites into two sublattices: those with $x_i+y_i+z_i$ odd (sublattice 1) and those with $x_i+y_i+z_i$ even (sublattice 2). Notice that two NN sites belong to different sublattices. In one sublattice sampling move, we choose one of the sublattices at random and then update the state of each site by computing the interaction with its NNs, $u_i({\hat\alpha})$, for the three orientations: ${\hat \alpha} = \hat{x}, \hat{y}, \hat{z}$. The new configuration is obtained by choosing, for each particle, a new orientation with probability: $p_i({\hat \alpha}) \propto \exp \left[ - \beta u_i({\hat \alpha}) \right]$.

In order to check the cluster algorithm and its implementation we have run pairs of simulations using either cluster moves or sublattices moves only. The results were found to be the same within error bars. The cluster algorithm is much more efficient than the sublattice algorithm and the relative efficiency increases as the system size $L$ increases.
Nevertheless, as we will discuss later, its performance is far from optimal for very large systems and temperatures slightly below the order-disorder transition.

The Monte Carlo simulations of the 3D model are run in cycles. We choose at random, with equal probability, one of the five  cycles to be run, namely (two) sublattice and (three) cluster samplings and then proceed as described above.

\section{Simulation results for the 3D SARR model}

\begin{figure}
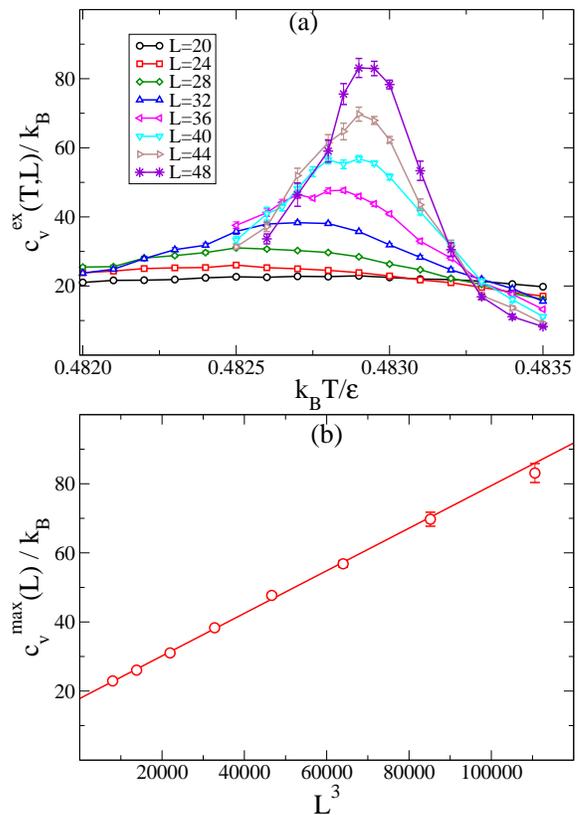

\includegraphics[width=75mm,clip=]{Fig1a.eps}
\includegraphics[width=75mm,clip=]{Fig1b.eps}
\caption{ (a) Excess heat capacities as a function of the temperature for different
system sizes (three-dimensional SARR model). (b)
Scaling of the peaks of the excess heat capacities with the system size.}
\label{FIG1}
\end{figure}

The order of the transition can be inferred from the scaling with the system size,
of the peak of the excess heat capacity:
$ c_{v}^{ex} = \left( \partial u  / \partial T\right)$, where $u \equiv U/L^3$ is the potential
energy per site.
At first order transitions the peak is expected to scale
as: \cite{Landau-Binder}
\begin{equation}
c_{v}^{\max}(L) = c_0 + \frac{ (\Delta U)^2}{4 k_B T_c^2 } L^3,
\end{equation}
where $T_c$ is the transition temperature.
In Figure \ref{FIG1} we plot the excess heat capacities in the transition region, and the scaling 
behavior of their peaks,
$c_v^{max}(L)$.
The scaling of $c_v^{max}(L)$ with $L^3$ indicates that the transition is first order.
A least-square fit yields the latent heat of the transition:
$\Delta U /(N\epsilon) = 0.024 \pm 0.001$ .

\begin{figure}
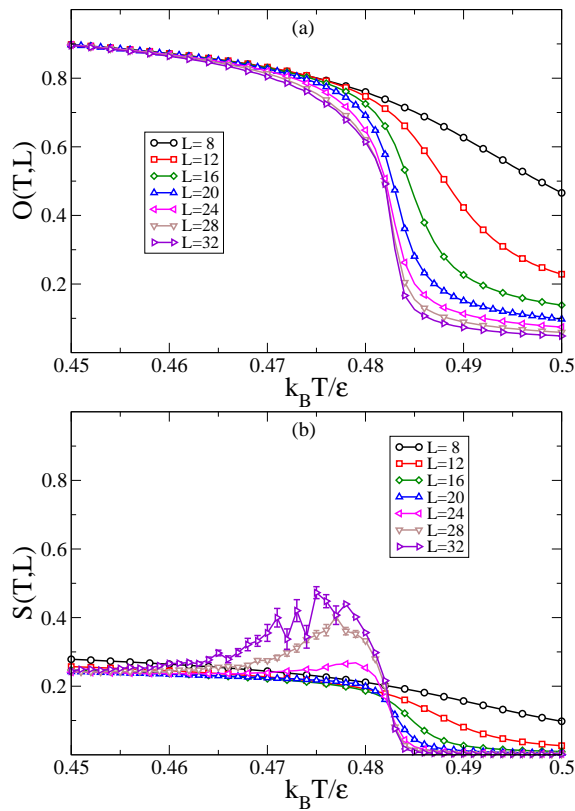

\includegraphics[width=75mm,clip=]{Fig2a.eps}
\includegraphics[width=75mm,clip=]{Fig2b.eps}
\caption{(a) Order parameter $O$, and (b) Order parameter $S$ for the three dimensional SARR model as
a function of the temperature for several system sizes, $L$ (see the legends).}
\label{FIG2}
\end{figure}

We consider two order parameters to characterize the transition.
The first: $O$ is based on the partial ground state order:
\begin{equation}
O = 1 -3 \min \left[ N_x, N_y, N_z \right]/N,
\end{equation}
where $N_{\alpha}$ is the number of sites with orientation $\alpha$, and $N=L^3$
is the total number of sites. The second: $S$ measures the uniaxial order:
\begin{equation}
S = \frac{1}{2} \left[ 3 \frac{ N_x^2 + N_y^2 + N_z^2 }{N^2} - 1 \right].
\end{equation}
In the ground state $O=1$ (one direction suppressed), while $<S> = 1/4$. Values of 
$S$ greater than $1/4$ below the order-disorder transition temperature signal the tendency for uniaxial order.
In Figure \ref{FIG2} we plot the two order parameters as a function of the temperature for several system sizes.
Both sets of curves $O(T,L)$and $S(T,L)$ show, as $L$ increases, an abrupt change at (or close to) the temperature where the heat capacity peaks. In small systems, $S(T)$ varies monotonically with the temperature. However,
for the largest systems $S(T)$ peaks at temperatures slightly below
the transition temperature, and saturates at values larger than the ground state value at low temperatures.
This finding suggests a surprising tendency for uniaxial order, i.e. at finite subcritical temperatures and large system sizes the particles prefer to align in one direction rather than aligning in two directions as expected from the ground state analysis. Notice that as $L$ increases the results for $S(T)$ below $T_c^*=k_BT_c/\epsilon  \simeq 0.483$ have large error bars. This is a signature
of the lost of efficiency of  the cluster algorithm to sample $S(T,L)$ slightly below $T_c$ as
the system size grows.

With the current algorithms and computational resources, however, we cannot establish the nature of the ordered phase at finite temperatures, in the thermodynamic limit.  

\section{Confined SARR model}

In order to investigate the mechanism that may drive uniaxial order in the 3D SARR model, and to quantify it, we have considered a simpler model where the system consists of a number of layers ($h$) with $L^2$ sites. These layers are taken perpendicular to $z$ direction. 
Periodic boundary conditions are only considered in directions $x$ and $y$.
The ground state of the confined SARR model is $2^h$-degenerate with all the particles in a given layer aligned along the $x$ or the $y$ direction. We note that geometrical confinement is also used to assist the self-assembly process of 3D systems and thus the study of confinement is also of some practical relevance \cite{Kretzschmar}. 

The confined systems are simulated using the algorithms described in section \ref{sec:sim}, with
minor adaptations: namely, the cluster moves are carried out only for layers perpendicular to ${\hat z}$.

The confined SARR model exhibits an order-disorder transition as the models in 2D and 3D. Note that the limit of confinement (single layer) of the 3D model is not equivalent to the 2D SARR model as the patches can be aligned in three distinct directions. Given the nature of the ground state, where each layer is ordered in an arbitrary direction ($x$ or $y$) we consider the single-layer order parameters $S_i$, defined as:
\begin{equation}
S_i= \frac{1}{L^2} \left[ N_x(i) - N_y(i) \right],
\end{equation}
where the index $i$ refers to one layer. $N_x(i)$ and $N_y(i)$ are the number of sites on layer $i$ with patches in the $x$ and $y$ directions, respectively.
Due to the symmetry of the model, the average single-layer order parameters vanish, $\langle S_i \rangle = 0$. In the thermodynamic limit, at low temperatures, we expect an ordering transition described by:
\begin{equation}
\langle  S_i^2 \rangle \left\{ \begin{array}{cc}  = 0; & \;\; T \ge T_c,  \\
  > 0 ; & \;\;  T < T_c. \end{array} \right.
\end{equation}
We anticipate a discontinuous transition for systems with a large number of layers (as in 3D), and a continuous one for thin slabs (as in 2D). The continuous transition is expected to be in the 2D Ising class as in single layer systems.

\section{Simulation Results for the confined systems}
\label{sec:results_confined}

In order to proceed we consider the scaling behavior of the heat capacity, or the related quantity $u'_{\beta}  \equiv (\partial u/\partial \beta)$, with $u$ the potential energy per site. For systems in the 2D Ising universality class, the scaling behavior at criticality is:
\begin{equation}
u'_{\beta} (\beta_c) \sim \ln L.
\end{equation}
In addition we investigate the scaling behavior of the ratios $g_{4i} = 
\langle S_i^4\rangle/\langle S_i^2\rangle^2$, related to the Binder cummulants \cite{Landau-Binder}. For 2D Ising critical behavior $g_{4}(L,\beta)$ for different $L$ cross at the universal value $g_{4} ^{(c)}\simeq 1.168 $ \cite{Salas}. 

\begin{figure}
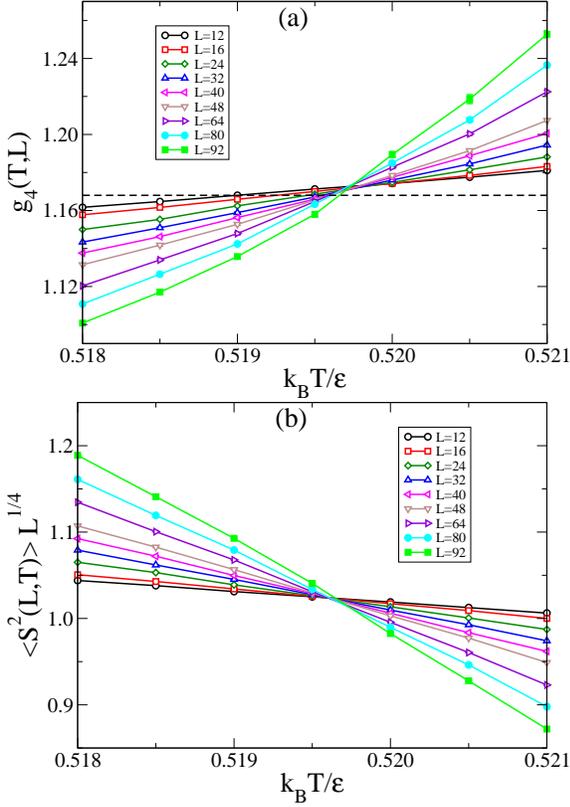

\includegraphics[width=75mm,clip=]{Fig3a.eps}
\includegraphics[width=75mm,clip=]{Fig3b.eps}
\caption{(a) Scaling behavior of $g_{4}(T)$, and (b) Scaling behavior of the order parameter,
$S^2$ for one-layer system systems ($h=1)$ and
different system sizes. The dashed line in (a) marks the universal value $g_4^{(c)}$ for
2D Ising critical behavior. The crossings of the curves $g_{4}(T)$ and
$<S^2(L,T)> L^{1/4}$ for different system sizes confirm
the expected 2D-Ising criticality.}
\label{FIG3}
\end{figure}

The results for one layer, $h=1$ comply with the expected 2D Ising critical behavior.
The critical temperature, $T_c$, is estimated from the Binder cummulant following standard procedures\cite{Almarza2010}. Considering system sizes in the range $12 \le L \le 96$ we find $T_c^*= k_B T_c/\epsilon = 0.5196 \pm 0.0001$.
This result is consistent with the behavior of the pseudocritical temperatures $T_c(L)$ defined by the peaks of the 
heat capacity as a function of $L$ (results not shown).
In addition, at the estimated $T_c$ the scaling of the average order parameter exhibits the expected Ising behavior: $S^2(L,T_c) \propto L^{-2 \beta'}$ (where $\beta'=1/8$ is the critical exponent for the magnetization). The results for $g_4(L,T)$ and the scaling of $S^2(L,T)$ are shown in Fig. \ref{FIG3}.

\begin{figure}
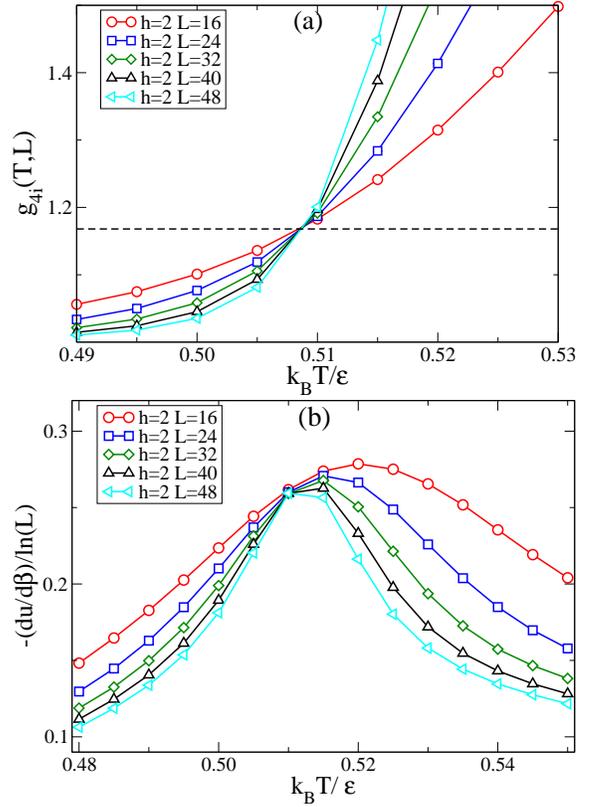

\includegraphics[width=75mm,clip=]{Fig4a.eps}
\includegraphics[width=75mm,clip=]{Fig4b.eps}
\caption{(a) $g_{4i}(T,L)$ for different system sizes, and
 (b) scaling of $(\partial u/\partial \beta)$ for the confined model with $h=2$. The system
size dependence of the results   suggest 2D Ising criticality (see the text).}
\label{FIG4}
\end{figure}

For $h=2$ and $h=3$, the curves for different system sizes
($ 12 \le L \le 48$) cross at a value of $g_{4i}$ close to that of the Ising universality class
(See Fig.\ref{FIG4}). Somewhat surprisingly, for $h=3$, the crossing of $g_{4i}(L,T)$ of the inner layer occurs at a temperature slightly below that of the outer layers  (plot not shown).
This is likely to be a finite-size effect. A possible explanation is that, for these values of $L$, the different layers are almost independent; within this assumption the {\it pseudo-critical} temperature of each layer depends on the density of defects
(number of sites with orientation ${\hat z}$). The inner layer is expected to have a larger number of defects since these sites oriented along ${\hat z}$ can establish two bonds; by contrast, in the outer layers the sites oriented along $z$ can form at most one bond. The larger density of defects reduces the stability of the ordered layer and thus its {\it local pseudo-critical temperature} is lower.

At finite temperatures, some particles will be aligned in the $z$ direction. An interaction between adjacent layers results from bond formation between particles in different layers (z-bonds) 
and a correlation between $S_{i}$ and $S_{i+1}$ may appear. If present, these correlations may drive the bias to uniaxial behavior observed in the simulations of the 3D model. 
 Let us define a global order parameter $S$ as:
\begin{equation}
S = \frac{1}{h} \sum_{i=1}^h S_i;
\end{equation}
where $h$ is the number of layers of the confined model. Again symmetry implies $\langle S \rangle=0$. The average value of $S^2$ may be written as:
\begin{equation}
\langle S^2\rangle = \frac{1}{h^2} \left[ \sum_{i}^h \langle S_i^2\rangle + 2\sum_{i=1}^{h-1} \sum_{j=i+1}^h \langle S_i S_j \rangle \right],
\label{ssquared}
\end{equation}
The correlation between two layers is defined as:
\begin{equation}
c_{ij} = \frac{ \langle S_i S_j \rangle }{ \left[ \langle S_i^2\rangle \langle S_j^2 \rangle \right]^{1/2} }.
\end{equation}
The correlation depends both on $L$ and $T$, $c_{ij}(L,T)$, and it is expected to vanish at low and high temperatures. 
Inspection of Eq.(\ref{ssquared}) reveals that $\langle S^2\rangle$ behaves in the limit  of  low temperatures as:
\begin{equation}
\lim_{T \rightarrow 0 } \langle S^2(L,T) \rangle = \frac{1}{h}.
\end{equation}

\begin{figure}
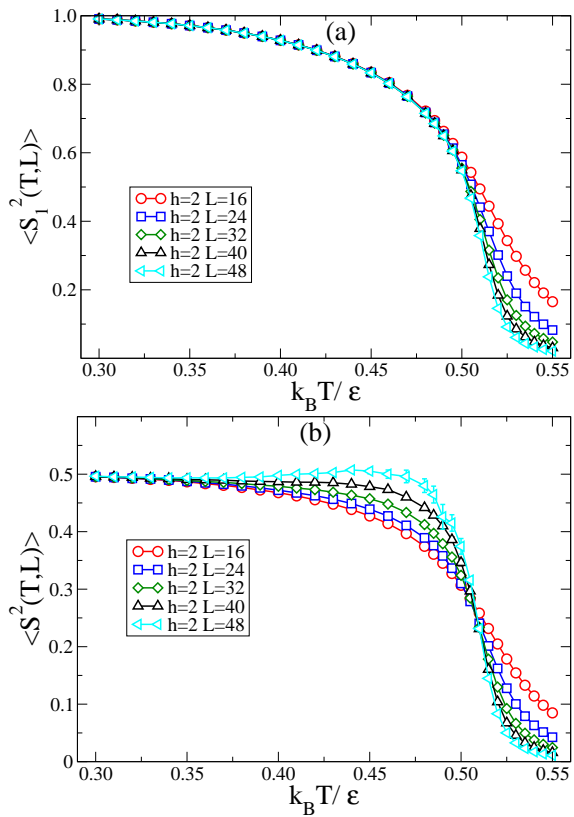

\includegraphics[width=75mm,clip=]{Fig5a.eps}
\includegraphics[width=75mm,clip=]{Fig5b.eps}
\caption{(a) Single layer order parameter for the confined model with $h=2$. (b) Global order parameter for the confined model with $h=2$.}
\label{FIG5}
\end{figure}

\begin{figure}
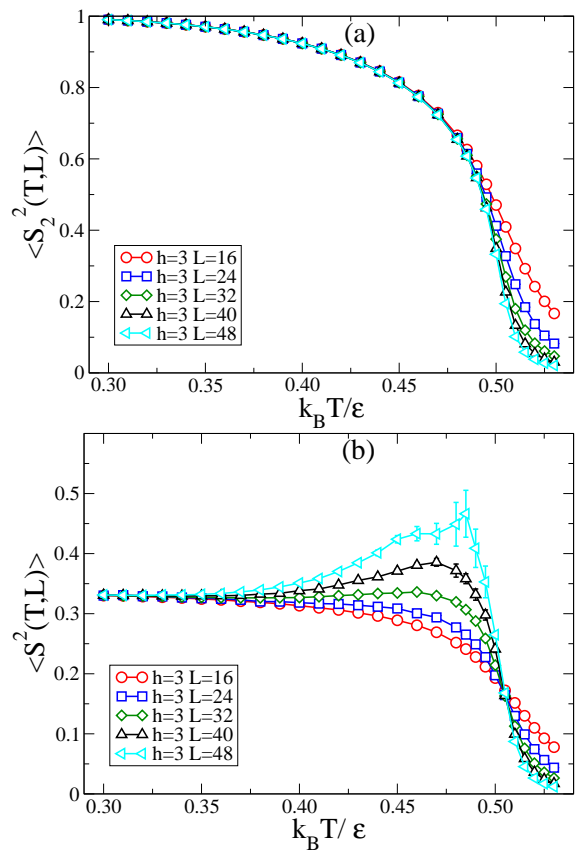

\includegraphics[width=75mm,clip=]{Fig6a.eps}
\includegraphics[width=75mm,clip=]{Fig6b.eps}
\caption{(a) Single layer (z=2),  and (b) global order parameters for the confined model with $h=3$}
\label{FIG6}
\end{figure}

We located the order-disorder transition of the model with $h=2$ in the full lattice limit by considering the behavior of $(\partial u/\partial \beta)$. In Figure \ref{FIG4}
we plot the results for different system sizes. In the region around the maximum we observe the scaling $(\partial u/\partial \beta) \sim \ln L$ in line with 2D Ising criticality.
In Figure \ref{FIG5} we plot the results for the single-layer, $S^2_1$(L,T), and the global, $S^2(L,T)$, order parameters for the same model. The single-layer order parameter exhibits the usual dependence on $T$ and $L$. The global order parameter, however, exhibits a different behavior: the curves for different system sizes cross around $T_c$, and then 
merge as the temperature decreases. 
The
 same qualitative behavior is observed for the confined model with $h=3$ in Figure \ref{FIG6}.
While the scaling of $(\partial u/\partial \beta)$ and $g_{4i}$ suggests a continuous transition, the crossing of the curves $<S^2>(L,T)$ at criticality suggests a (weak) first-order transition, as the order parameter, $|S|$, in the
thermodynamic limit, could exhibit  a discontinuity at the transition jumping from zero $(T > T_c)$ to a finite value $|S_c|>0$.

\begin{figure}
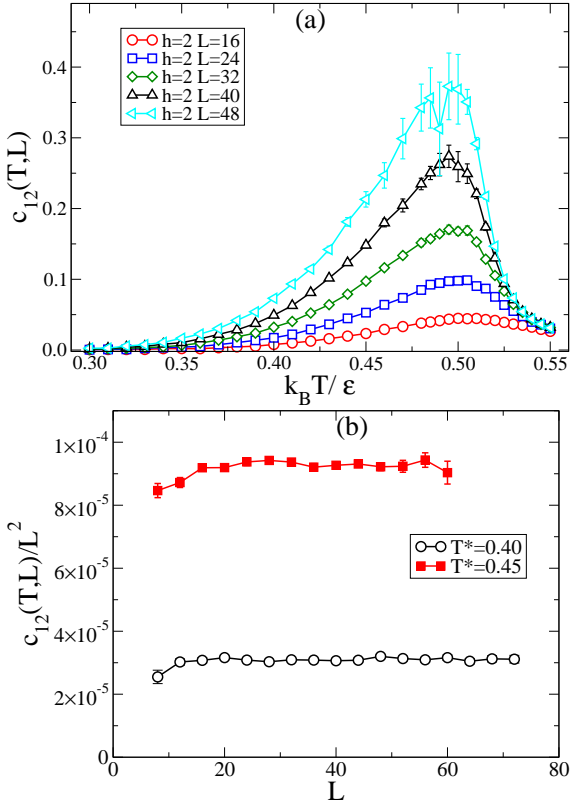

\includegraphics[width=75mm,clip=]{Fig7a.eps}
\includegraphics[width=75mm,clip=]{Fig7b.eps}
\caption{(a) Correlation function, $c_{12}$ between the order parameters of adjacent layers, $h=2$.
(b) Dependence of $c_{12}$ on the lateral size of the systems with $h=2$ at two temperatures.}
\label{FIG7}
\end{figure}

In Figure \ref{FIG7}a we plot the correlation function $c_{12}$ between the layers of the $h=2$ system. As expected the correlation decreases and appears to vanish at low and high temperatures. The most relevant feature, however, is that the correlation between layers increases markedly with the system size, $L$. Figure \ref{FIG7}b reveals that the correlation increases as $L^2$ (in the range of sizes considered). These results suggest that for $T>0$, in the thermodynamic limit $L \rightarrow \infty$, the confined model becomes uniaxial (i.e., the layers will align along a unique direction).

\begin{figure}
\includegraphics[width=75mm,clip=]{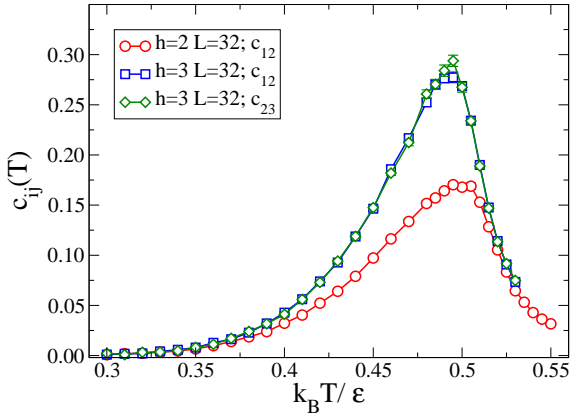}
\caption{Correlation functions between the order parameters of adjacent layers for $h=2$ and
$h=3$, (with lateral system size $L=32$).}
\label{FIG8}
\end{figure}

Now, we consider the effect of the number of layers on the correlation between adjacent layers. In Figure \ref{FIG8} we plot the layer-layer correlation $c_{ij}$ for $h=2$ and $h=3$, for systems with $L=32$. Note that the correlation functions $c_{12}$ and $c_{23}$ are equal (except for statistical errors) due to the symmetry of the model. The main conclusion from the results of Fig \ref{FIG8} is that for a fixed value of $L$ the correlation between adjacent layers increases with the number of layers. This suggests that the ordered phase of the three dimensional SARR model may become uniaxial, in the thermodynamic limit, at finite temperatures $0 < T < T_c$. Note, however, that as the temperature decreases the system size required to observe uniaxial ordering increases very rapidly.

\begin{figure}
\includegraphics[width=75mm,clip=]{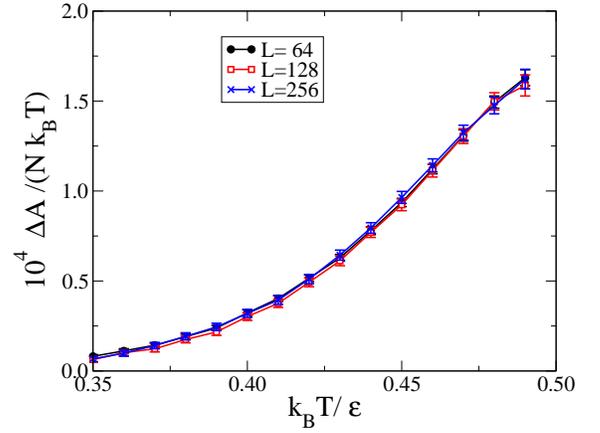}
\caption{Free energy difference of the configurations of two layer systems where the layers are oriented in distinct 
and in the same directions.}
\label{FIG9}
\end{figure}

It is clear that the simulation algorithms used in this work loose efficiency 
as $L$ increases at temperatures slightly below the critical temperature. In order to confirm the
trend to uniaxiality suggested by the results presented so far, we return to the two layer
system, and use an indirect method to compute the free energy difference $\Delta A = A_{xy}-A_{xx}$,
where the subscripts indicate configurations where the layers are oriented preferentially in the 
same ($xx$) and in different $(xy)$ directions.
$\Delta A$ is computed for large $L$ using thermodynamic integration from low temperature (where the free 
energy of the two types of configurations is the same: $\Delta A(T_0) = 0$, as $T_0 \rightarrow 0$). 
The free energy difference is then:
\begin{equation}
\frac{\Delta A(T)}{T} = \frac{\Delta A(T_0)}{T_0}  + \int_{T_0}^T \Delta U(T') d \frac{1}{T'}
\end{equation}
where $\Delta U(T) = U_{xy}(T)-U_{xx}(T)$. The potential energies $U_{\alpha \beta}(T)$
are computed using Monte Carlo simulation of relatively large systems, $L=64$, $L=128$ and $L=256$, without cluster moves
to avoid interconversion between the two types of configurations.
The results are plotted in Figure \ref{FIG9}. $\Delta A$ increases with temperature and is proportional to $L^2$. Thus 
for large systems and moderate temperatures, the confined SARR model is expected to exhibit uniaxial order. 

\section{Discussion}

\begin{figure}
\includegraphics[width=75mm,clip=]{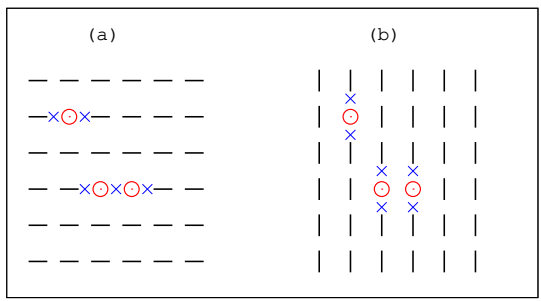}
\caption{Sketch of the effect of NN pairs of $z$-sites (or $z$-bonds) in the orientational correlation of the layers. Segments represent particles with patches aligned in the plane of the layer, circles represent sites with patches aligned in the $z$ direction. Crosses mark the in-plane bonds suppressed by the presence of $z-$sites. Note that the number of such 
bonds depends on the alignment of the pair of $z$-sites with respect to the alignment within the layer. Therefore the configuration (a) has a lower energy than the configuration (b).}
\label{FIG10}
\end{figure}

These surprising results may be interpreted as follows: Consider a two-layer system at low temperature with most particles aligned along the $x$ or $y$ directions. At $T>0$, however, a number of particles will align along $z$. A $z$-bond lowers the energy by $-\epsilon$ with
respect to two independent $z$-sites, one in each layer, but isolated $z$-bonds do not contribute to the orientational correlation between layers. Now, suppose that two $z$-bonds occur in NN positions (for instance one between sites $r_{a1}=(i,j,1)$ and $r_{a2}=(i,j,2)$, and the second between sites $r_{b1}=(i+1,j,1)$ and $r_{b2}=(i+1,j,2)$) (See Figure \ref{FIG10}). At (low) subcritical temperatures this pair of NN $z$-bonds promotes the alignment of both layers in the direction defined by the pair (the $x$ direction in this example as shown in Figure \ref{FIG10}. In practice, configurations with a different number of pairs of NN $z-$bonds along the $x$ and $y$ directions favor the alignment of the layers.
A bond counting argument gives the probability of aligning one layer along the easy $x$ direction over the probability of aligning it along $y$, when a single pair of NN $z$-bonds, along $x$, is present:

\begin{equation}
\frac{p_x}{p_y} =  e^{\beta \epsilon};
\end{equation}
The ratio of the probabilities of aligning the layers over the probability of not doing so is then:
\begin{equation}
\frac{p_{xx} + p_{yy} } { p_{xy} + p_{yx} } = \cosh ( \beta \epsilon ).
\end{equation}  
An estimate of the density of NN $z$-bonds, $\alpha(T)$ is:
\begin{equation}
\alpha (\beta) \approx e^{-4\beta \epsilon}.
\end{equation}
Note that only configurations where the number of NN $z$-bonds in the $x$ and $y$ directions are different contribute to the orientational correlation of the layers. Let us, however, consider the rough estimates given above. At $T^*=0.40$, $\beta \epsilon  =2.5$ and $\alpha(T) \approx  4.5 \times 10^{-5}$. For a system with $L=64$ most configurations will not have NN $z$-bonds, and about one in five (0.186) will have one. An estimate of $c_{12}$ is then:
\begin{equation}
c_{12} \simeq 0.186 \times \frac{ \cosh(2.5)  - 1 }{ \cosh(2.5) + 1 } \simeq   0.134.
\end{equation}
which is close to the value obtained from the simulation $c_{12} = 0.125\pm 0.007$. This estimate supports the hypothesis that the uniaxial behavior results from the orientational correlation between adjacent layers driven by the presence of NN $z$-bonds. 

The characterization of the ordering transition of the confined SARR model with $h\ge 2$ requires the development of more efficient cluster simulation algorithms and thus the behavior of the 3D SARR model cannot be investigated at present. The problem is related, but not identical, to the model for crystallization and vitrification of semiflexible living polymers investigated by Menon and co-workers in 2D and 3D \cite{MenonEPL,MenonPRE}. 

\acknowledgments
We acknowledge M. Sim\~oes for stimulating discussions in various stages of this work.
NGA gratefully acknowledges the support from the Direcci\'on General de Investigaci\'on Cient\'{\i}fica  y T\'ecnica under Grant No. FIS2010-15502, and from the
Direcci\'on General de Universidades e Investigaci\'on de la Comunidad de Madrid under Grant No. S2009/ESP-1691 and Program MODELICO-CM. MMTG and JMT acknowledge 
financial support from the Portuguese Foundation for Science and Technology (FCT) under Contracts nos. PEst-OE/FIS/UI0618/2011 and PTDC/FIS/098254/2008.

\end{document}